\documentclass[12pt,preprint]{aastex}

\usepackage{subfigure}

\slugcomment{to appear in the {\it Astronomical Journal}}

\shorttitle{Parallaxes from CTIOPI}
\shortauthors{Jao et al.}

\begin{document}

\title{The Solar Neighborhood XXIV. Parallax Results from the CTIOPI
  0.9-m Program: Stars with $\mu$ $\ge$ 1\farcs0 yr$^{-1}$ (MOTION
  Sample) and Subdwarfs}

\author{Wei-Chun Jao\altaffilmark{1}, Todd J. Henry\altaffilmark{1}}
\affil{Department of Physics and Astronomy, Georgia State University,
 Atlanta, GA 30302}
\email{jao@chara.gsu.edu, thenry@chara.gsu.edu}

\author{John P. Subasavage\altaffilmark{1}}
\affil{Cerro Tololo Inter-American Observatory, La Serena, Chile}
\email{jsubasavage@ctio.noao.edu}  

\author{Jennifer G. Winters\altaffilmark{1}, Adric
R. Riedel\altaffilmark{1}}

\affil{Department of Physics and Astronomy, Georgia State University,
Atlanta, GA 30302}
\email{winters@chara.gsu.edu, riedel@chara.gsu.edu}

\and 

\author{Philip A. Ianna\altaffilmark{1}}
\affil{Department of Astronomy, University of Virginia,
Charlottesville, VA 22904-4325}
\email{philianna3@gmail.com}

\altaffiltext{1}{Visiting Astronomer, Cerro Tololo Inter-American
Observatory.  CTIO is operated by AURA, Inc.\ under contract to the
National Science Foundation.}

\begin{abstract}

We present 41 trigonometric parallaxes of 37 stellar systems, most of
which have proper motions greater than 1\farcs0 yr$^{-1}$.  These are
the first trigonometric parallaxes for 24 systems.  Overall, there are
15 red dwarf systems and 22 red subdwarf systems in the sample.  Five
of the systems are multiples with directly detected companions, and we
have discovered perturbations caused by unseen companions in two
additional cases, the dwarf LHS 501 and the subdwarf LHS 440. The
latter system may eventually provide important dynamical mass points
on the subdwarf mass-luminosity relation.  Two additional stars of
note are LHS 272, the third closest M-type subdwarf at a distance of
only 13.6 pc, and LHS 2734AB, a high velocity subdwarf binary with
$V_{tan}>$ 700 km/sec, which likely exceeds the escape velocity of the
Milky Way.  We also report the first long term variability study of
cool subdwarfs indicating that cool subdwarfs are less photometrically
variable than their main sequence counterparts.

\end{abstract}

\keywords{astrometry --- solar neighborhood --- stars: distances ---
stars: late-type --- subdwarfs}

\section{Introduction}

This is the sixth list of trigonometric parallaxes (hereafter,
$\pi_{trig}$) from the Cerro Tololo Inter-American Observatory
Parallax Investigation (CTIOPI) using data from the CTIO 0.9-m
telescope.  Previous papers reported parallaxes of stars from the
MOTION sample \citep[stellar systems with $\mu$$\geq$1\farcs0
yr$^{-1}$,][]{Jao2005}, new members of the RECONS 10 pc sample
\citep{Henry2006}, a young brown dwarf, \citep[2MASSW
J1207334$-$393254,][]{Gizis2007}, white dwarfs \citep{Subasavage2009}
and the SLOWMO sample \citep[stellar systems with
0\farcs5$\leq\mu<$1\farcs0 yr$^{-1}$,][]{Riedel2010}. In this paper,
we target both MOTION stars and subdwarfs, presenting new $\pi_{trig}$
for 24 stellar systems and improved $\pi_{trig}$ for 13 additional
systems.  Such high proper motion stars are prime targets for parallax
studies because they may be nearby --- useful for luminosity and mass
function studies, as well as being the closest representatives of
their types --- or they may be rare subdwarf members in the solar
vicinity with high velocities.  The overlap in these two samples makes
it natural to combine the two types of objects in this paper.
Twenty-nine of the systems are from the MOTION sample and are split
into 14 red dwarf and 15 subdwarf systems. The remaining eight systems
include seven additional subdwarfs and a red dwarf, LHS 3740 with a
$\mu$ of nearly 1\farcs0 yr$^{-1}$.

\section{Observations and Data Reduction}

\subsection{Astrometry}
\label{sec:astrometry}

The CTIO 0.9-m telescope has a 2048$\times$2048 Tektronix CCD camera
with 0\farcs401 pixel$^{-1}$ plate scale \citep{Jao2003}. For both
astrometric and photometric observations, we use the center quarter of
the chip, yielding a 6\arcmin8 square field of view. Parallax frames
are taken through one of four filters, $V_J(old)$, $V_{J}(new)$,
$R_{KC}$ or $I_{KC}$\footnote{The central wavelengths for the
$V_{J}(old)$, $V_{J}(new)$, $R_{KC}$ and $I_{KC}$ filters are 5438,
5475, 6425, and 8075 \AA, respectively. The old and new $V$ filters
are discussed in next paragraph.}  (hereafter without the subscripts),
so that either science or reference stars have maximum peak counts of
$\sim$50,000 (saturation occurs at 65,535 counts) for better
centroiding.  The magnitudes of CTIOPI targets are 9 $\leq$ $VRI$
$\leq$ 19.  Depending on the brightness of the science targets,
reference stars, and sky conditions, exposure times vary from 20 to
1200 seconds.  With few exceptions, observations are made within
$\pm$30 minutes of a science target's transit to minimize the
corrections required for differential color refraction.  Typically,
three to 10 frames are taken in each night, depending primarily on the
exposure time required.  Bias and dome flat frames are taken nightly
to enable routine calibration of the science images.

One ``event'' during the CTIOPI effort warrants special attention
here.  We have used two $V$ filters, dubbed the ``old'' Tek\#2 $V$
filter ($\lambda_{central}=5438$\AA, $\Delta\lambda=1026$\AA) and
``new'' Tek\#1 $V$ filter ($\lambda_{central}=5475$\AA,
$\Delta\lambda=1000$\AA), during the 11 years of observations because
the ``old'' filter cracked in February 2005.  The ``new'' $V$ filter
was used between 2005 and 2009. In July 2009, the old $V$ filter was
reinstated for use after confirming that the crack in the corner did
not significantly affect astrometric residuals.  We have found that as
long as at least 1--2 years of data (depending on observing frequency)
have been taken in both filters, the data can be combined to determine
a reliable $\pi_{trig}$ \citep{Subasavage2009}. In total, 10 of the 37
systems discussed in this paper were observed astrometrically in the
$V$ filter. Further details about the filters and their effects on the
astrometry can be found in \citet{Subasavage2009} and
\citet{Riedel2010}.

The stellar paths traced by science stars on the sky are combinations
of proper motions and parallactic shifts. \citet{Jao2005} and
\citet{Henry2006} include extensive discussions of the data reduction
processes used to separate these two movements.  Briefly, we (1) use
SExtractor \citep{sextractor} to measure centroids, (2) define a
six-constant plate model to find plate constants (given in equation 4
of \citealt[]{Jao2005}), (3) assume that ensembles of reference stars
have zero mean parallax and proper motion, (4) solve least-square
equations for multi-epoch images (given in equation 5 of
\citealt[]{Jao2005}), and (5) convert from relative parallax to
absolute parallax by estimating the distance of the reference stars
photometrically.  The correction from relative to absolute parallax is
accomplished using photometric distance estimates by comparing $VRI$
colors to $M_{V}$ for single, main sequence, stars in the RECONS 10 pc
sample \citep{Henry1997, Henry2004}.  For each reference star a
distance is estimated, and the correction to absolute parallax is then
computed using the weighted mean distance of the entire reference
field.  The error on the correction is determined using equation (6)
in \citet{Jao2005}.

\subsection{Photometry}

We have obtained $VRI$ photometry of these targets at the CTIO 0.9-m
telescope using the same instrumental setup used for the astrometry
frames. Similar to those astrometry observations, bias and dome flat
frames are taken nightly for basic image reduction.  Most science
stars were observed at $\sec z$ $<$ 1.8 or less (a few were observed
between 1.8 and 2.0 airmasses because of extreme northern or southern
declinations).  Various exposure times were used to reach S/N $>$ 100
for science stars in each of the $VRI$ filters.  Combinations of
fields that provided 10 or more standard stars from
\citet{Landolt1992} and/or E-regions from \citet{Graham1982} were
observed several times each night to derive transformation equations
and extinction curves. Further details of photometric data reduction,
definition of transformation equations, errors, etc.~can be found in
\citet{Jao2005} and \citet{Winters2011}.

\subsection{Spectroscopy}

Spectroscopic observations were made using the CTIO 1.5-m with the R-C
spectrograph and Loral 1200$\times$800 CCD camera.  Grating \#32 was
used in first order with a tilt of 15.1$^\circ$, and observations were
made using a 2\arcsec~slit.  The order-blocking filter OG570 was
utilized to provide spectra covering the range of 6000\AA~to
9500\AA~with a resolution of 8.6\AA.  Bias frames, dome flats, and sky
flats were taken at the beginning of each night for calibration.

At least two exposures were taken for each object to permit cosmic ray
rejection, with additional exposures taken if stars were particularly
faint.  A 10 second Ne$+$He$+$Ar or Ne only arc lamp spectrum was
recorded after each target to permit wavelength calibration.  Several
spectroscopic flux standard stars found in the $IRAF$ spectroscopy
reduction packages were observed during each observing run, usually
nightly.  Reductions were carried out using $IRAF$ reduction packages
--- in particular {\it onedspec.dispcor} for wavelength calibrations
and {\it onedspec.calibrate} for flux calibrations.  We use the same
set of spectral standard dwarfs discussed in \cite{Jao2008} to assign
spectral types for dwarfs, and follow the same method discussed in
that paper to assign spectral types for subdwarfs.

\section{Astrometric Results}
\label{sec:results}

Parallax results for the 37 systems are given in Table 1. The single
measurement accuracy for well-balanced reference fields with exposures
at least a few minutes in duration is typically 2--8 mas.  For weak
reference fields (fewer than five reference stars, lopsided reference
star configurations, or very faint reference stars), shorter exposure
times, and close binaries with asymmetric point spread functions, the
single measurement accuracy may be as high as 20 mas. Ultimately, the
final absolute parallax errors for this sample of stars are 0.9--3.5
mas.

In Table 1, we present details about the astrometric observations
(filters used, number of seasons observed, number of frames used in
reductions, time coverage, span of time, and the number of reference
stars) and results (relative parallaxes, parallax corrections,
absolute parallaxes, proper motions, position angles of the proper
motions, and the derived tangential velocities based on relative
proper motions given in column 12). Twenty-four of the 37 systems
discussed here had no previous $\pi_{trig}$.  These systems are listed
in the top portion of the table. The remaining 13 systems shown in the
bottom of Table 1 have $\pi_{trig}$ previously reported in \citet{YPC}
(hereafter YPC), \citet{Jao2005}, \citet{Smart2007}, or
\citet{Bartlett2009}. Figure~\ref{fig:color.mag.1} shows the HR
diagram of these targets and we will discuss individual systems later
in Section~\ref{sec:notes}. The astrometric, photometric, and
spectroscopic results presented here supersede those in
\citet{Jao2005} because additional data and improved reduction
techniques have been used, as discussed in detail in
\citet{Subasavage2009}.

\section{$VRI$ Photometry} 

New $VRI$ photometry for the 37 systems is given in
Table~\ref{tbl:phot.result}, as well as the near-infrared photometry
($J$, $H$, and $K_{s}$ bands) from the Two Micron All Sky Survey
\citep{2mass}.  Names are given in the first two columns, followed by
the optical $VRI$ photometry in columns 3--5, the number of nights
(column 6) on which $VRI$ observations were taken.  The information
used for the photometric variability (see Section~\ref{sec:varia}) is
listed in columns 7--9.  The $JHK_{s}$ photometry is given in columns
10--12. All of the systems have at least two nights of $VRI$
photometry data.  The mean errors of $VRI$ photometry are usually
$\leq$ 0.03 mag.  However, stars that are faint at $V$ sometimes have
larger errors, e.g.~the faintest star, SIPS 1529$-$2907 with
$V$=19.38, has the largest S/N error of 0.12 mag.  The remainder of
these parallax targets have S/N error less than 0.07 mag at $V$.  See
\citet{Winters2011} for a detailed discussion of our $VRI$ photometry
program errors.

The $V$ photometry given here represents combined values from the two
different $V$ filters discussed in Section~\ref{sec:astrometry}.  To
check the consistency of photometry from these two filters, we
selected ten photometric standard stars observed between 2001 and 2009
on a total of 45 photometric nights, including 20 nights of photometry
in the ``old'' $V$ filter and 25 nights of photometry in the ``new''
$V$ filter.  The ten stars selected were SA98-670 ($V=$11.93),
SA98-650 ($V$=12.27), SA98-676 ($V$=13.07), SA98-671 ($V$=13.39),
SA98-675 ($V$=13.40), and SA98-682 ($V$=13.75) from
\cite{Landolt1992}, E2-o ($V$=14.09), E2-t ($V$=14.60), and E2-I
($V$=15.17) from \cite{Graham1982} and GJ 406 ($V$=13.57) from
\cite{Kilkenny1998}.  The $V-I$ colors of these standard stars are
between 0.20 and 4.12.

\begin{figure}
  \includegraphics[angle=90, scale=0.7]{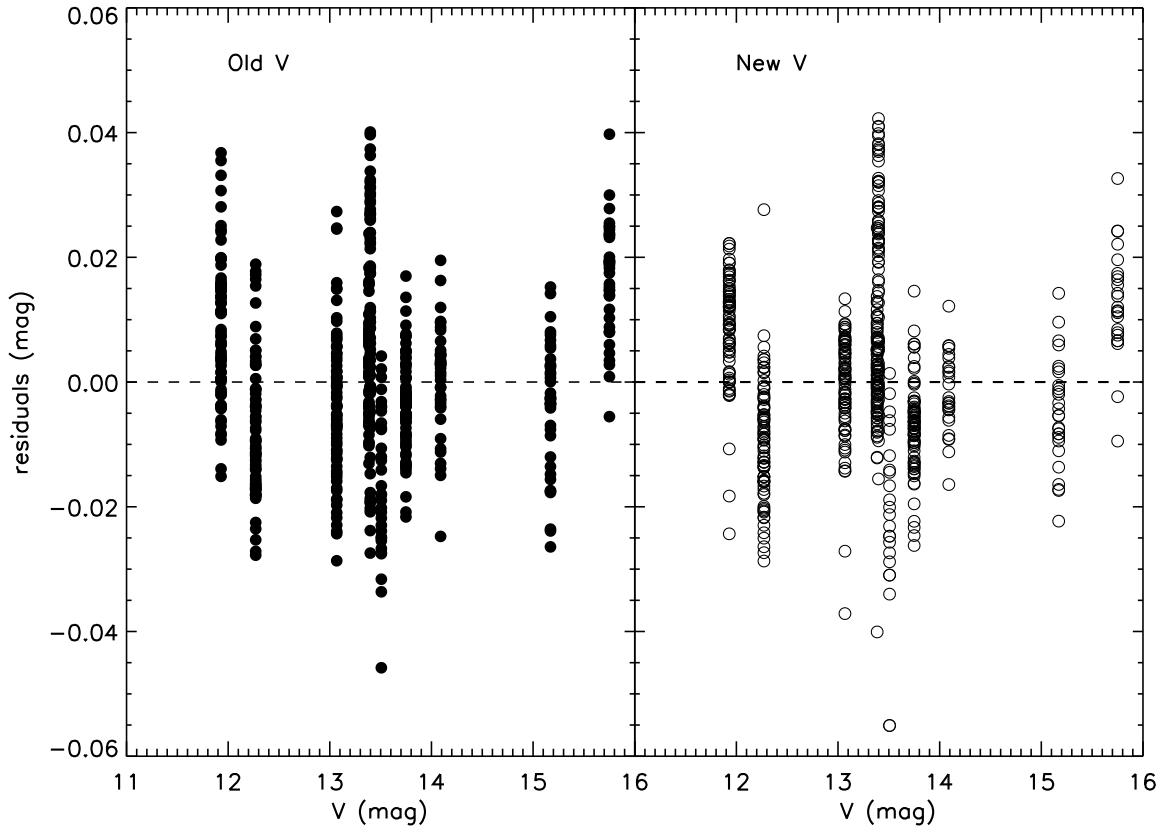}
  
  \caption{Differences between reported magnitudes and our derived
  magnitudes are shown for ten photometric standard stars observed
  between 2001 and 2009 through two $V$ filters.  Left (solid
  circles) and right (open circles) panels indicate stars observed
  with ``old'' and ``new'' $V$ filters, respectively.  The dashed
  line indicates identical magnitudes derived from our measurements
  when compared to magnitudes in the photometric standard papers.
  Note that SA98-671 ($V$=13.39) and SA98-675 ($V$=13.40) overlap in
  this figure because of the scale of these plots.  There are no
  obvious systematic differences seen for the two filters.}

  \label{fig:standardV}
  \end{figure}

The differences between the measured $V$ magnitudes for the standards
relative to their quoted magnitudes in the original photometric
standards papers are plotted in Figure~\ref{fig:standardV}.  It is
clear that the two filters yield effectively identical $V$ band
photometry for these ten standard stars.  The mean differences are
0.006 and 0.007 mag from the ``old'' and ``new'' $V$ filters,
respectively, which is much less than our quoted 0.03 mag error.
Hence, we conclude that the $V$ band photometry from the two filters
can be combined.

\section{Variability Studies}
\label{sec:varia}
\subsection{Analysis}
\label{sec:varia.analysis}

In addition to the $VRI$ photometry, we take advantage of our
long-term observations of these fields (up to 10 years) in the
parallax filter (column 7 in Table~\ref{tbl:phot.result}) to analyze
the photometric variability (column 8 in Table~\ref{tbl:phot.result})
of these parallax targets. The total number of nights used for the
variability study for each star is given in column 9 in
Table~\ref{tbl:phot.result}.

Here we examine 130 red dwarfs for evidence of variability, as
measured by the standard deviations, $\sigma_{mag}$, of magnitudes
derived relative to reference stars in the fields.  Combining the 41
stars in this paper with stars in \citet{Henry2006} and
\citet{Riedel2010}, two previous papers with similar data streams,
yields a total of 135 stars.  We exclude five stars from the study, as
follows.  GJ 1207, with $\sigma_{mag}=$ 0.263 \citep{Henry2006}, was
observed during an obvious flare in a single series of observations.
LP 771$-$095BC, with $\sigma_{mag}=$ 0.043 \citep{Henry2006},
exhibited a sudden $\sim$0.3 magnitude drop in brightness at $V$ on UT
1999 August 21, but LP 771$-$095A, the primary star, and none of the
reference stars showed this event (one possible cause is an eclipse).
LHS 193B, with $\sigma_{mag}=$ 0.049 (this work), is a white dwarf
that was underexposed in all frames because the primary, LHS 193A, was
the star targeted for parallax.  Similarly, LHS 2734A was the parallax
target, so LHS 2734B with $\sigma_{mag}=$ 0.020 (this work) was
systematically underexposed and has been omitted from the analysis.
Finally, as discussed below in $\S$\ref{sec:varia.results}, the
photometry for LHS 272 with $\sigma_{mag}=$ 0.015 (this work) was
corrupted by a background galaxy.

To evaluate variability, we first extracted the instrumental
magnitudes using SExtractor with a fixed 6\arcsec~diameter
aperture. Because most parallax frames were taken on nights when
photometric standard stars were not observed, we do not convert
instrumental magnitudes to apparent magnitudes; we instead compare the
relative fluxes of the target stars to the reference stars used to set
the astrometric grid.  We follow the methodology discussed in
\citet{Honeycutt1992} to handle an inhomogeneous set of exposures.
The basic equation of condition used to evaluate inhomogeneous sets of
exposures is

\begin{equation}
m_{i}^{j}=m0_{i}+\delta m^{j},
\end{equation}

\noindent where $m_{i}^{j}$ is the instrumental magnitude of star $i$
on image $j$ (magnitude output from SExtractor), $m0_{i}$ is the mean
instrumental magnitude of star $i$, and $\delta m^{j}$ is the
magnitude offset for image $j$. In other words, $\delta m^{j}$ takes
into account the differences caused by sky conditions, extinction, and
exposure times for each frame in the series.  We then utilize a
Gaussfit program \citep{Jefferys1987} to perform least-square
calculations to minimize the deviations between the two sides of this
equation of condition, as described in more detail in
\citet{Honeycutt1992}.

After obtaining $\delta m^{j}$, the ``corrected'' instrumental
magnitude ($mc_{i}^{j}$) is

\begin{equation}
mc_{i}^{j}=m_{i}^{j}-\delta m^{j}.
\end{equation}

\begin{figure}
  \centering
  \includegraphics[angle=90, scale=0.55]{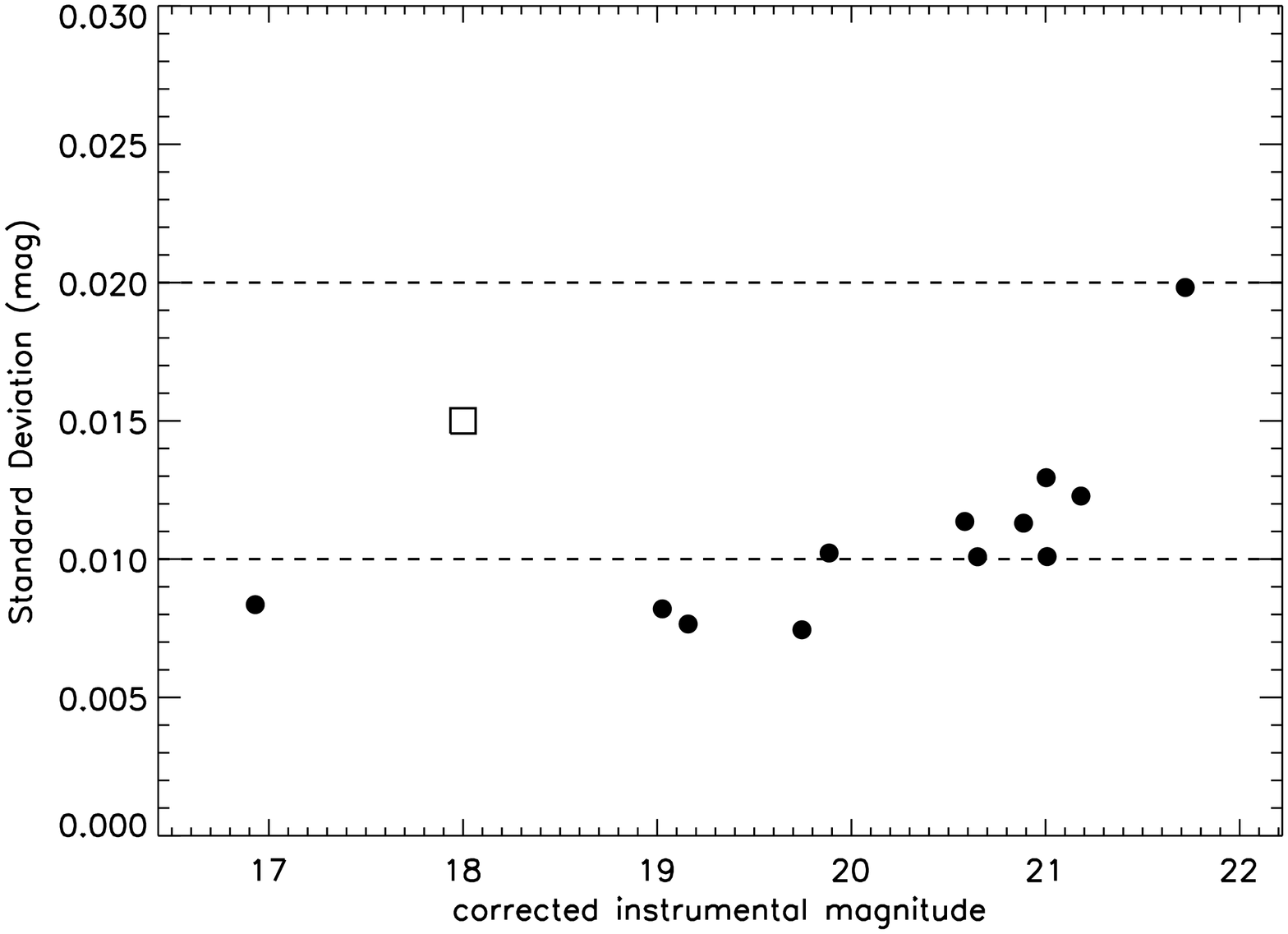}
  \includegraphics[angle=90, scale=0.55]{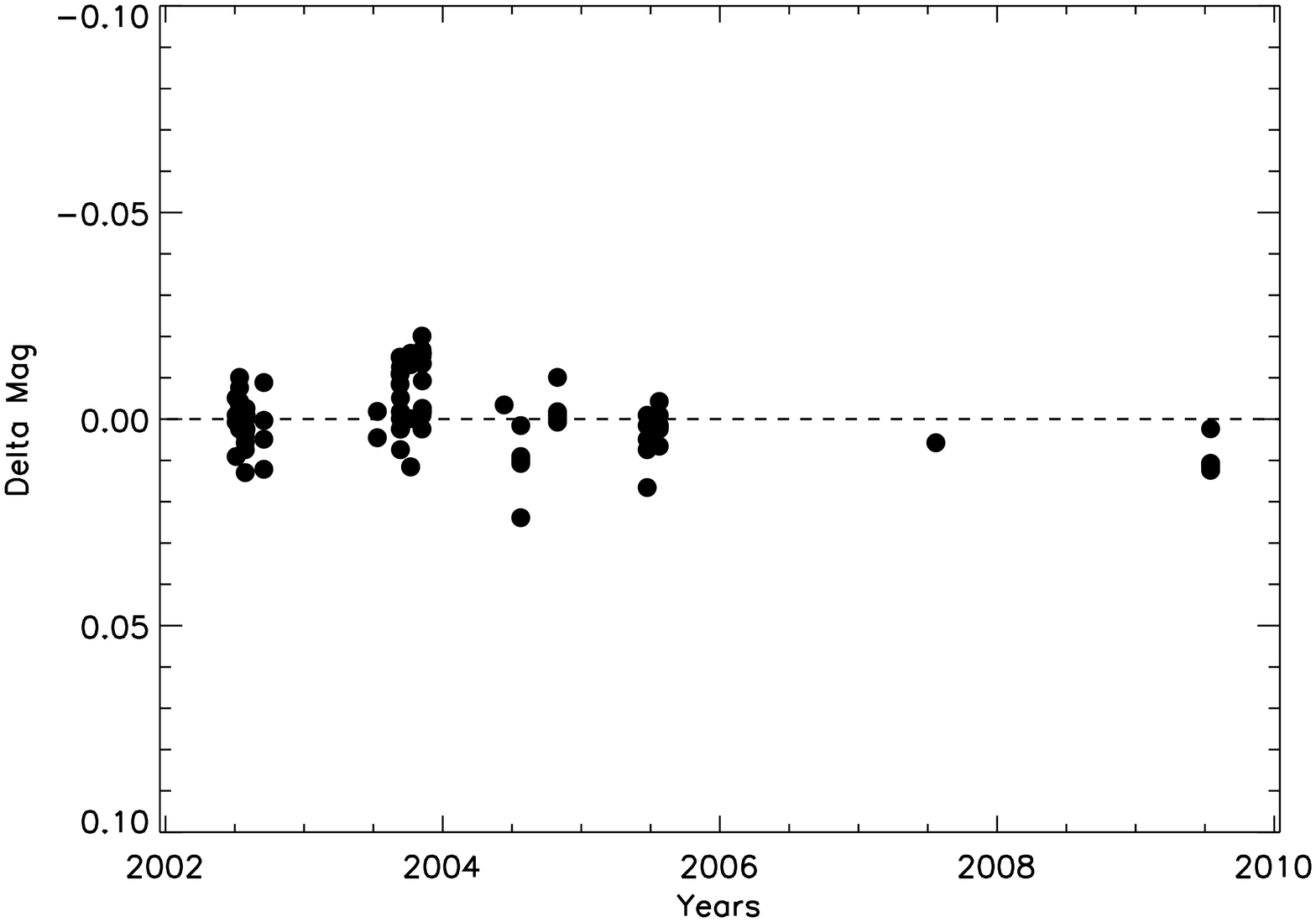}
  
  \caption{A variability study of LHS 518. Top: This plot shows
  corrected instrumental magnitudes, ($mc_{i}^{j}$), in the $R$ band
  and standard deviations, $\sigma_{mag}$, of 12 stars in the field of
  LHS 518.  The brightest point at the far left represents LHS 518.
  An open box simulates an outlier star with unusually high standard
  deviation that would be discarded before deriving the final
  $\sigma_{mag}$ value of LHS 518.  Bottom: The light curve of LHS 518
  is shown around its mean magnitude. Its standard deviation (0.008
  mag) is calculated using all of the images shown from 18 nights of
  observations.}
  
  \label{fig:varia.sep}
  \end{figure}

\noindent We then calculate the $\sigma_{mag}$ values of the corrected
instrumental magnitudes, $mc_{i}^{j}$, in a given filter for all
science and parallax reference stars in that field. An example output
is shown in the top of Figure~\ref{fig:varia.sep}.  Naturally, the
standard deviation is a function of magnitude, where fainter stars
have larger standard deviations.  In column 8 of
Table~\ref{tbl:phot.result}, we list the standard deviations in
magnitudes for the science stars.  A high standard deviation value for
a science or reference star can be caused by (1) a flare occurring
during a series of observations on one night, (2) long term
variability, (3) contaminated photometry from a nearby source, (4)
high sky background caused by moon illumination or a twilight
sky\footnote{In order to get high parallax factors during astrometric
measurements, we sometimes take frames during evening and morning
nautical twilight when the center of the Sun is about 11 degrees below
the horizon, with consequently high sky background.}, or (5) faint
reference stars in the field.  Only the first two causes are
indicative of true variability.

If a reference star's magnitude appears to have a high standard
deviation, as illustrated with an open box shown in the top panel of
Figure~\ref{fig:varia.sep}, we remove this outlier and perform another
round of least-square calculations to obtain the $\delta m^{j}$ values
for the frames and the final $mc_{i}^{j}$ and derived $\sigma_{mag}$
values.  Such outliers can be caused by any of the five reasons listed
above.  Usually, after the second round of calculations to get
$mc_{i}^{j}$, we determine the final magnitudes for the science star
for each frame, which represents its variability around its mean
magnitude, as shown in the bottom of Figure~\ref{fig:varia.sep}.

We have carried out three specific tests to support the variability
results, including analyses of (1) the two $V$ filters used, (2)
reference star selection, and (3) the target star brightness ranking
relative to the reference stars.

{\it Two $V$ Filters:} To check that slight differences between the
two $V$ filters (``old'' and ``new'') do not cause significant extra
variability for those parallax stars observed using both filters, we
have evaluated the photometric time series of three white dwarfs, LHS
145, GJ 440 and GJ 781.3, observed during CTIOPI
\citep{Subasavage2009}.  These three stars have rich datasets taken
using both filters and are known to be non-pulsating white dwarfs,
which ensures extremely low levels of intrinsic variability.  The
results given in Table~\ref{tbl:varia.wd} show that in both filters,
our ``observational floor'' for detecting variability is $\sim$0.008
mag (discussed further in $\S$5.2). Thus, the use of different V
filters may introduce small additional uncertainities. However, the
"observational floor" in V stays below 0.010 mag.

\begin{figure}
  \centering
 
  \voffset-20pt{
  \includegraphics[angle=90, scale=0.7]{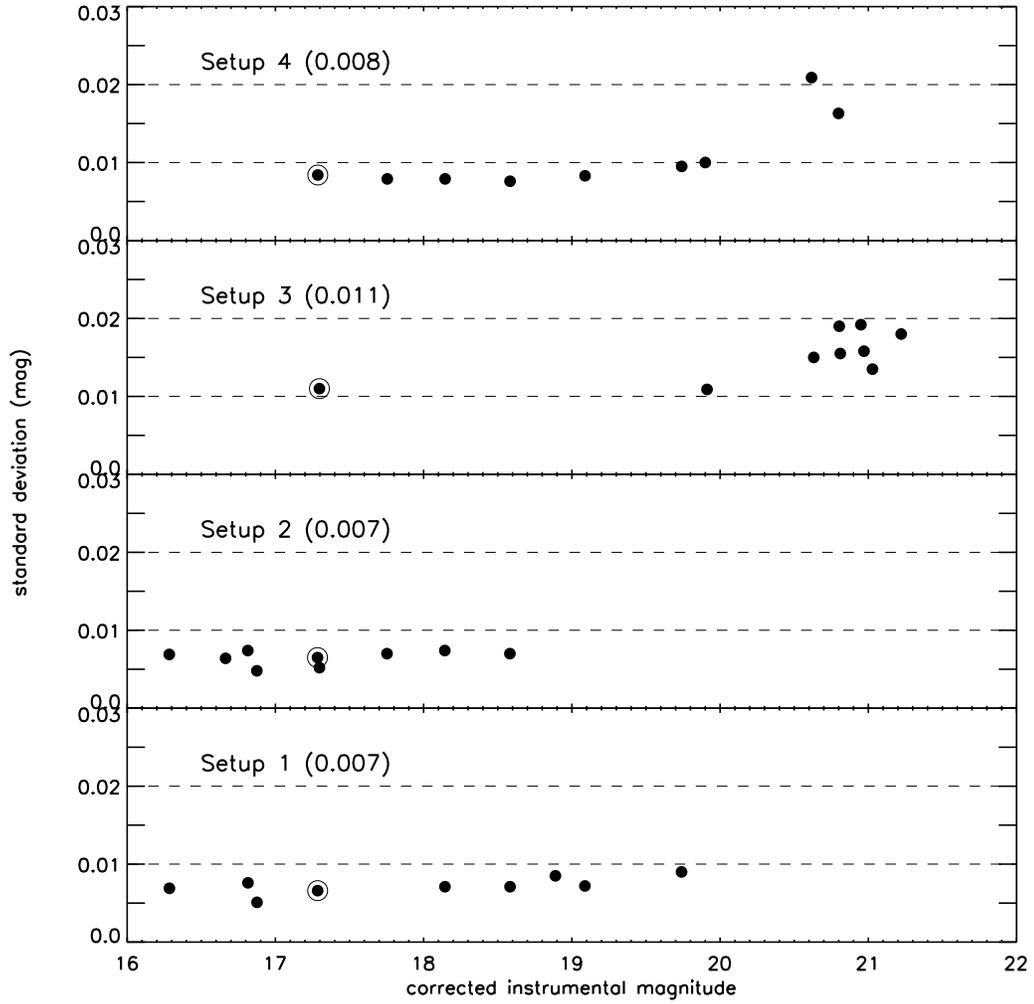}
  }
  \caption{Standard deviations of LHS 3045 (given in the parenthesis)
  from different sets of reference stars in the field are shown.
  Points surrounded by open circles indicate LHS 3045 in each
  configuration.  The details of different setups are discussed in
  $\S$5.1.}
  \label{fig:lhs3045}
  \end{figure}

{\it Reference Star Selection:} To investigate whether or not the
reference stars' brightnesses affect $\delta$$m^{j}$, and consequently
our measurements of variability, we have selected the reference star
rich field of LHS 3045 for a series of tests in which we keep the
total number of reference stars constant.  Figure~\ref{fig:lhs3045}
shows results using four different sets of reference stars.  Setup 1
is the one used for the parallax reduction reported in
Table~\ref{tbl:pi.result} and the variability result given in
Table~\ref{tbl:phot.result}.  Setup 2 includes three stars that differ
from those in Setup 1 so that the mean magnitude of the Setup 2
reference stars is brighter than for Setup 1.  Setup 3 includes a set
of faint reference stars that could be selected in this field (but
wouldn't, given much more suitable reference stars).  Setup 4 also
includes reference stars that are all fainter than LHS 3045, but which
are generally brighter than in the case of Setup 3.

Setups 1 and 2 have identical standard deviations for all stars in
common.  Setup 2 was not chosen for the parallax work because some
stars are on the edge of the CCD chip, where their PSFs are slightly
distorted and degrade the astrometric results.  Predictably, Setup 3
yields the highest standard deviations among the tests because all of
the reference stars are fainter than mag 19.8.  In practice, these
stars cannot be selected for astrometric work because they are too
faint to determine reliable centroids if the target star is kept
unsaturated.  As discussed in $\S$2.1, to derive good centroids for
astrometric measurements, we expose long enough to let either parallax
or reference stars reach peak counts of $\sim$50,000 for centroiding,
which consequently provides a high S/N for all stars used in the
astrometric reduction.  In fact, because the target/reference star
configuration used in Setup 3 will not provide a reliable parallax, we
would drop this target from the parallax program.  We did not select
Setup 4 because suitable brighter stars were available for both
astrometric and photometric work.  The $\sigma_{mag}$ values from
Setup 4 are 0.001 mag higher than for Setups 1 and 2 because of four
stars of mag 19.7 or fainter.  Nonetheless, the $\sigma_{mag}$ value
for the target star remains less than 0.010 mag and is only 0.001 mag
higher than for Setups 1 and 2.  We therefore conclude that all
realistic reference star configurations yield variability measurements
consistent at the 0.001 mag level.

\begin{figure}
  \centering
  \hoffset-20pt{
  \includegraphics[angle=90, scale=0.85]{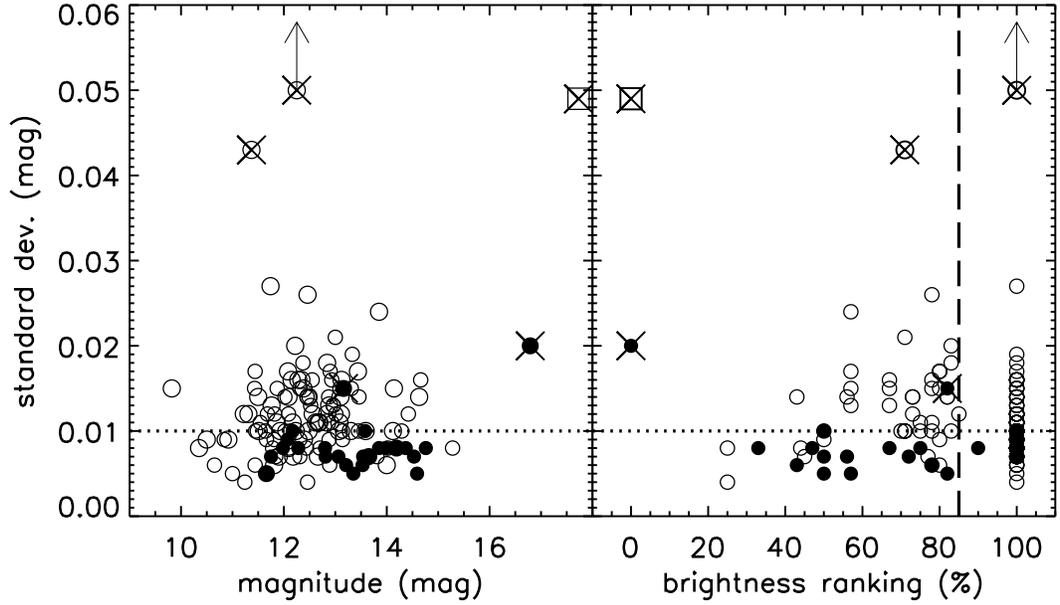}
  }
  \caption{Variability measurements of 108 dwarfs (open circles) and
  22 subdwarfs (solid circles) are plotted against their apparent
  magnitudes (left panel) and brightness rankings among all stars in
  each field (right panel).  In the left panel, the apparent
  magnitudes correspond to $V$, $R$ or $I$ values, depending on the
  filter used for parallax observations.  In the right panel, a
  vertical dashed line shows the mean brightness ranking of $\sim$85\%
  for all 130 stars.  Dotted lines mark a standard deviation value,
  $\sigma_{mag}$, of 0.010 mag.  Five stars discussed in
  $\S$\ref{sec:varia.analysis} are marked with Xs. The white dwarf,
  LHS 193B, is represented as an open box.}

  \label{fig:varia1}
  \end{figure}

{\it Target Star Brightness:} To better understand how variability
measurements change with the brightness ranking of the target star, in
Figure~\ref{fig:varia1} we plot the $\sigma_{mag}$ values
vs.~magnitude for the 130 sample stars, and show the brightness
ranking of the target stars in the right panel.  A science star with a
brightness ranking of 85\% means that 85\% of reference stars in that
field are fainter than the science star.  A science star with
brightness ranking of almost 100\% has a field similar to Setup 4 in
Figure~\ref{fig:lhs3045}.  The mean brightness ranking of all 130
stars is $\sim$85\%, marked as a vertical dashed line. The right panel
of Figure~\ref{fig:varia1} shows that there is no clear relation
between a science target's brightness ranking and its variability for
most stars, which are found in the 50--100\% region.

Overall, we conclude that none of the three tests indicates any
systematic effects on the variabilities measured for the target stars.

\subsection{Variability Results}
\label{sec:varia.results}

\begin{figure}
  \centering
  \includegraphics[angle=90, scale=0.8]{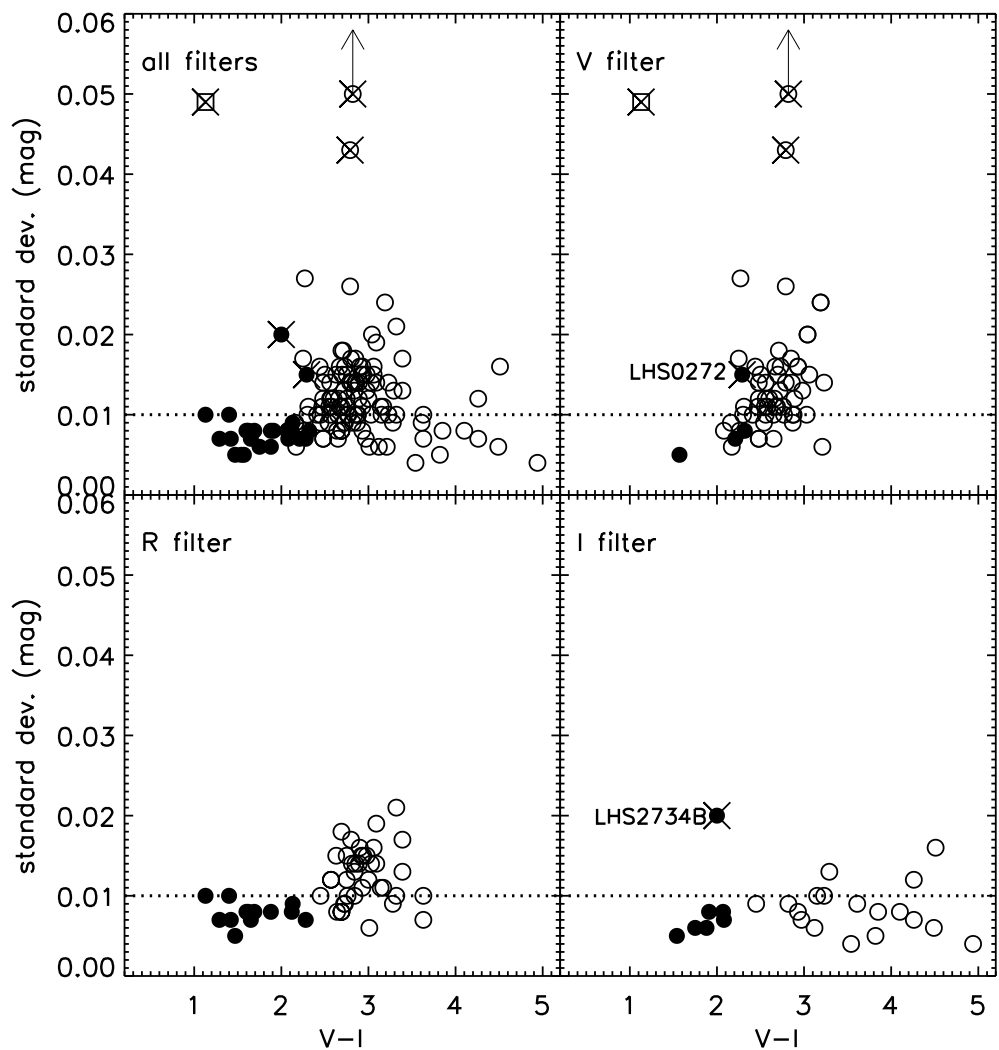}
  
  \caption{Variability measurements of 108 dwarfs (open circles) and
  22 subdwarfs (solid circles) are plotted against their $(V-I)$
  colors.  The entire sample is shown in the upper left panel.  Stars
  observed in $V$, $R$ and $I$ filters are plotted separately in the
  other three panels.  LHS 272 and LHS 2734B are labeled and discussed
  in the text. Symbols have the same meanings as in
  Figure~\ref{fig:varia1}.}
 
  \label{fig:varia2}
  \end{figure}

Figure~\ref{fig:varia2} shows stellar variability, measured using
$\sigma_{mag}$, as a function of color for the 130 dwarfs and
subdwarfs included in this study, separated into panels that include
all stars and split into subsets for stars observed in the $V$, $R$,
and $I$ filters.  Points with Xs are the five stars omitted from the
statistical study, for reasons given in $\S$5.1.  For the sample of
108 dwarfs (open circles in Figure~\ref{fig:varia2}), the mean
variabilities at $V$, $R$, and $I$, are 0.013 (51 stars), 0.013 (39
stars), and 0.008 (18 stars) mag, respectively.  These results
indicate that as a group, red dwarfs are more variable in the $V$ and
$R$ bands than in the $I$ band.  We also perform a Kolmogorov-Smirnov
test for 51 stars at $V$ band and 18 stars at I band. The result shows
a probability of 99.87\% that distributions of these two bands are
statistically different.

For the sample of 22 cool subdwarfs, the average variabilities are
0.007 at $V$ (3 stars), 0.008 at $R$ (13 stars), and 0.007 at $I$ (6
stars) mag, which are lower than for dwarfs at $V$ and $R$.  A
Kolmogorov-Smirnov test for the $R$ band sample, in which we have the
largest number of subdwarfs, indicates a probability of 100\% that
the dwarf and subdwarf cumulative distributions are different, where
the greatest absolute difference between the two distributions is
equal to 0.74, highly indicative of different samples.  Thus, it
appears that the variability in subdwarfs is lower than that of
dwarfs, at least in the $R$ band, implying that their atmospheres are
less active.  More subdwarfs need to be observed in the $V$ band to
confirm our suspicion of less activity than for subdwarfs than for
dwarfs at $V$, although the data do point in that direction.  We also
note that subdwarf variability is similar in all three filters, rather
than increasing at shorter wavelengths, which is the case for dwarfs.
This may be indicative of quiet atmospheres in the subdwarfs, for
which we have reached our ``observational floor'' (about 0.008 mag) in
our ability to measure variability in the current datasets, which
suspiciously also matches the value for the least variable $I$ filter
measurements for dwarfs. Although this comparison is based on
relatively small samples, it is the first time the long-term
variability of cool subdwarfs has been investigated\footnote{We note
  that there are numerous publications concerning the variability of
  hot ``subdwarfs'' because of their pulsations, but they are
  fundamentally different from the ``cool subdwarfs'' discussed
  here.}.

Overall, most subdwarfs' variability is less than 0.01 mag, with two
notable exceptions, both of which have been excluded in the
statistical evaluations of the subdwarf sample above:

{\bf LHS 272} This star is relatively bright ($V=$ 13.16), yet has the
largest $\sigma_{V}=$ 0.015 of the 11 stars measured in that field.
For comparison, a relatively faint ($V=$ 14.17) reference star has
$\sigma_{V}=$ 0.007.  However, LHS 272 was observed on the wing of a
bright background galaxy $\sim$17\arcsec~distant at a position angle
of 217$^{\degr}$ in 2001.  In our frames, LHS 272 has moved quickly
away at $\mu=1.439$ yr$^{-1}$ at 279.2$^{\degr}$, but the galaxy may
corrupt the variability measurements for those images taken during the
first few epochs of observations, causing an erroneously high
variability measurement.

{\bf LHS 2734B} This is the fainter star of a pair of science targets
in the field; the primary, LHS 2734A was targeted for parallax so the
B component was systematically underexposed, resulting in a high
variability measurement due to poor S/N, with $\sigma_{I}=$ 0.020 mag.
This is also the faintest star observed among all 130 systems, with
$I=$ 16.79.

So far, we have only observed a few dozen subdwarfs over these long
time periods, and with limited numbers of reference stars.  At
present, we are comparing subdwarfs observed at $R$ with $V-I<$ 2.4 to
dwarfs observed at $R$ with $V-I>$ 2.4, as shown in
Figure~\ref{fig:varia2}.  In order to make a systematic comparison of
dwarfs and subdwarfs of similar color, we need long-term photometric
monitoring campaigns that cover large swaths of sky, such as the
Hungarian-made Automated Telescope Network (HATNet) or the All Sky
Automated Survey (ASAS).  For example, \citet{Hartman2009} reported
variability for 27,560 K and M dwarfs using HATNet data, but this rich
dataset has not yet been applied to subdwarfs.  The catch is that
these surveys typically reach $\sigma_{mag}\sim$ 0.01 mag at $I\sim$
11, a magnitude limit too bright to reach many cool subdwarfs.  To
reach large numbers of faint, cool, subdwarfs, we will need the next
generation of survey projects like the Large Synoptic Survey Telescope
(LSST), the Panoramic Survey Telescope and Rapid Response System
(Pan-STARRS), and the SkyMapper project.

\section{Notes on Individual Systems}
\label{sec:notes}

\begin{figure}
  \includegraphics[angle=90, scale=0.7]{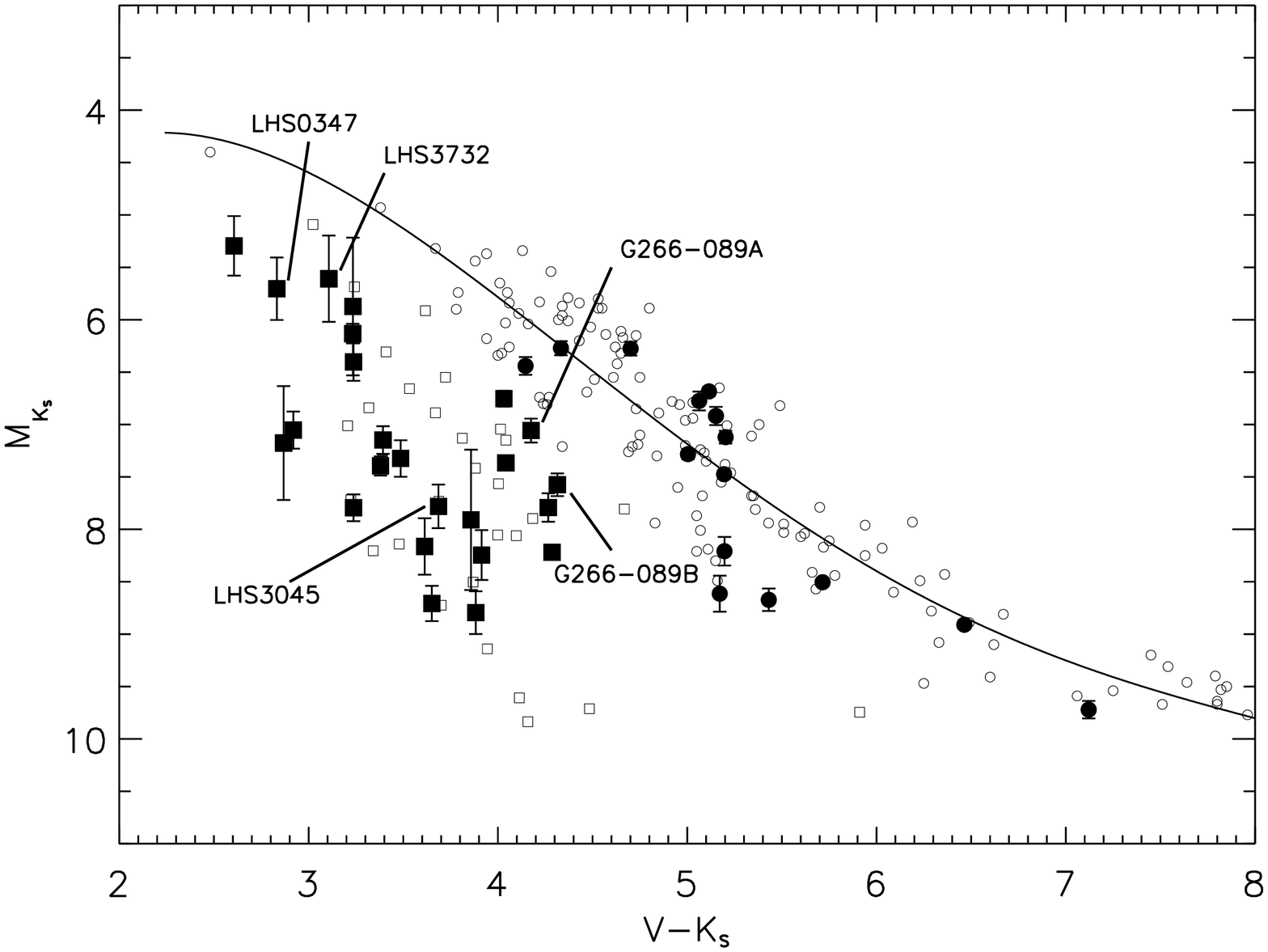}
  
  \caption{The HR diagram, using $M_{K_{s}}$ vs.~$V-K_{s}$, is shown
  for 41 stars in the 37 systems outlined in Tables 1 and 2.  Filled
  boxes and circles indicate subdwarfs and dwarfs, respectively. Open
  boxes represent 32 subdwarfs (LHS stars with $\mu >~$1\farcs0
  yr$^{-1}$) from \citet{Gizis1997}.  Open circles represent RECONS
  sample members ({\it www.recons.org}) and some very late M dwarfs
  discussed in \citet{Henry2004}, with an empirical fit tracing the
  main sequence stars.  Some stars discussed in Section~\ref{sec:notes}
  regarding their locations on the HR are labeled here.}
  
  \label{fig:color.mag.1}
  \end{figure}

{\bf G 266-089A and B} These stars comprise a resolved binary system
also known as LHS 111 (A) and LHS 110 (B).  The separation is
9\farcs02 at position angle 321.1$^{\circ}$ \citep{Jao2003}, a
separation that allows individual parallaxes to be determined for the
two components.  The weighted mean of the two measurements is
$\pi_{trig}=$ 33.75 $\pm$ 1.20 mas.  At a distance of 29.6 pc, this
measured separation corresponds to a projected separation of $\sim$270
AU, so we do not expect to see any orbital motion during the $\sim$5
years of astrometric observations.  \citet{Bidelman1985} reported
spectral types of M3.5V and M4.0V for A and B, respectively.  We
report a joint spectrum for the combined system because both stars
fell in the slit during acquisition, and the individual spectra were
not sufficiently separated on the chip for reliable individual
reductions.  The spectrum indicates dwarf features, but both stars
fall below the main sequence, as shown in
Figure~\ref{fig:color.mag.1}.  We therefore note this system as
M4.0J[VI] (J $=$ joint) until we obtain separate spectra.

{\bf LSR 0627$+$0616} A single image from the first epoch extends the
astrometric coverage from 2.97 years to 6.07 years.  The parallax
error drops from 1.33 to 1.25 mas, while the proper motion error drops
from 1.2 to 0.9 mas yr$^{-1}$.

{\bf LHS 272} With $\pi_{trig}$ $=$ 73.95 $\pm$ 1.18 mas (13.53 $\pm$
0.22 pc), this is the third closest M-type subdwarf found to date.
Only Kapteyn's star with $\pi_{trig}$ $=$ 255.27 $\pm$ 0.86 mas (3.92
$\pm$ 0.02 pc, weighted mean parallax from YPC and Hipparcos) and
$\mu$ Cas B with $\pi_{trig}=$ 132.57 $\pm$ 0.57 mas (7.54 $\pm$ 0.04
pc, weighted mean parallax from YPC and Hipparcos) are closer.  The
fourth closest known M subdwarf, LHS 20, is at 62.40$\pm$3.30 mas or
16.07 $\pm$ 0.85 pc (YPC). We do not detect a perturbation by an
unseen companion of LHS 272 over $\sim$4 years of data, nor do we see
any companions via optical speckle observations \citep{Jao2009}.

{\bf LHS 327} \citet{Augensen1978} commented that this star had a
total space velocity of 582 km sec$^{-1}$ and would escape from the
Galaxy. However, \citet{Ryan1991} reported $V_{rad}=$ 80 km/sec, from
which we derive a total space velocity of 430 km sec$^{-1}$. This star
may therefore be marginally bound to the Galaxy because the Galactic
escape velocity is estimated to be between 450 and 650 km/sec
\citep{Leonard1990}.

{\bf LHS 347} This is an early K type star and is below the main
sequence with a relatively large error bar, as shown in
Figure~\ref{fig:color.mag.1}. However, its spectrum is almost
identical to a K2.0V dwarf. Because of its high $V_{tan}=$ 331.3 km
sec$^{-1}$, and low luminosity, we assign it a type K2.0[VI],
representing its uncertain assignment as a subdwarf.

{\bf LHS 2734AB} This is a binary with separation 68\farcs8 at
position angle 162.3$^{\circ}$.  We find different relative
trigonometric parallaxes for LHS 2734 A (2.79 mas) and B ($-$1.70
mas).  The system is too far away for us to measure a meaningful
parallax, given that $\pi_{rel}^{A}/\sigma_{\pi_{rel}^{A}}\sim2.3$ and
$\pi_{rel}^{A}/\pi_{corr}^{A}\sim2.4$.  The secondary is 2.2 mag
fainter than the primary at $I$, the filter used for astrometric
measurements.  Thus, B is underexposed in the images, so we have
adopted A's absolute parallax for B in Table~\ref{tbl:pi.result}.
Using our new $\pi_{trig}$, we find that the system has $V_{tan}>$ 700
km/sec, although this value is highly uncertain.  Nevertheless,
$V_{tan}$ is certainly large, and this system may be able to escape
the Galaxy.

{\bf LHS 3045} This has astrometric residuals indicative of a possible
perturbation in the RA axis, but not in DEC axis. The perturbation is
not seen in reference star residuals, lending credence to its
veracity.  We will continue to observe the star to confirm or refute
the trend.  \citet{Bidelman1985} reported a spectral type of K5V, but
its location in Figure~\ref{fig:color.mag.1} shows that it is clearly
in the subdwarf region, and a [VI] spectral type is given in
Table~\ref{tbl:phot.result}. A spectroscopic observation is needed to
confirm its luminosity class.

{\bf LHS 3732} \citet{Ibata1997} report this star to be a subdwarf and
its position in Figure~\ref{fig:color.mag.1} ($(V-K)$ = 3.11)
indicates that it a subdwarf of early K type.  Our measured proper
motion of 0\farcs959/yr and parallax of 8.34 $\pm$ 1.58 mas yield
$V_{tan}\sim$550 km/sec.  \citet{Augensen1978} reported $V_{rad}=$ 120
km/sec, which results in $(U,V,W)$\footnote{This is a velocity
relative to the Local Standard of Rest (LSR), where the solar motion
is (U$_{\odot}$, V$_{\odot}$, W$_{\odot}$)=($+$11.1, $+$12.24,
$+$7.25) km sec$^{-1}$ with respect to the LSR \citep{Schonrich2010}.}
$\approx$ ($-$48,$-$522,$-$141) km sec$^{-1}$.  The total space
velocity is $\sim$543 km/sec, indicating that the star may not be
bound to the Galaxy.  We note that nearly all of the velocity
contribution is a result of the $V_{tan}$ measurement, which is based
on a large, and relatively poorly known, distance.

{\bf LHS 318} \citet{Smart2007} reported $\pi_{trig}$ $=$ 24.8$\pm$6.0
mas.  Our parallax of 18.76$\pm$2.32 mas is consistent with their
measurement, with a smaller error by a factor of 2.6.

{\bf LHS 326} \citet{Smart2007} reported $\pi_{trig}$ $=$ 11.7$\pm$4.3
mas.  Our parallax of 20.39$\pm$1.94 mas is marginally consistent with
their measurement, with a smaller error by a factor of 2.2.

\begin{figure}
  \centering \subfigure[]{ \includegraphics[scale=0.35,angle=90]{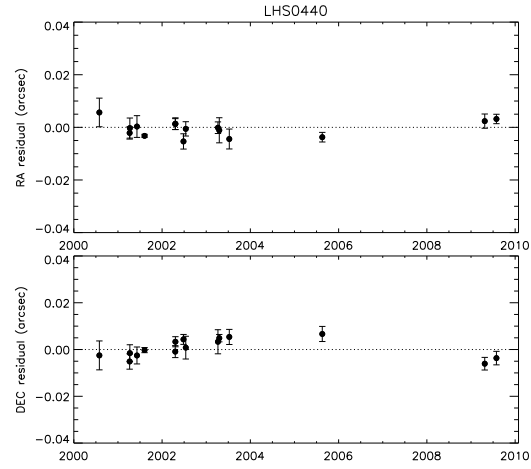}
  \label{fig:lhs0440}} 
  \subfigure[]{\includegraphics[scale=0.35,angle=90]{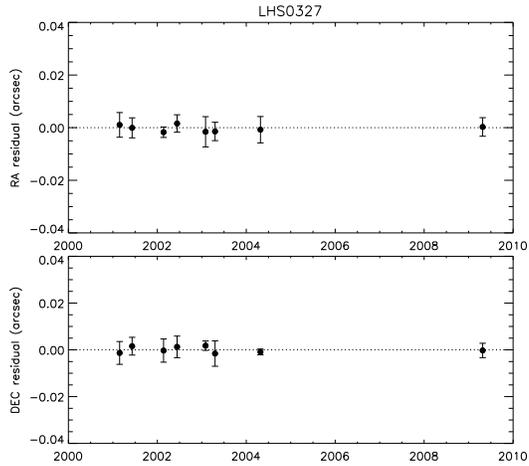}
  \label{fig:lhs0327}} 
  \subfigure[]{\includegraphics[scale=0.35,angle=90]{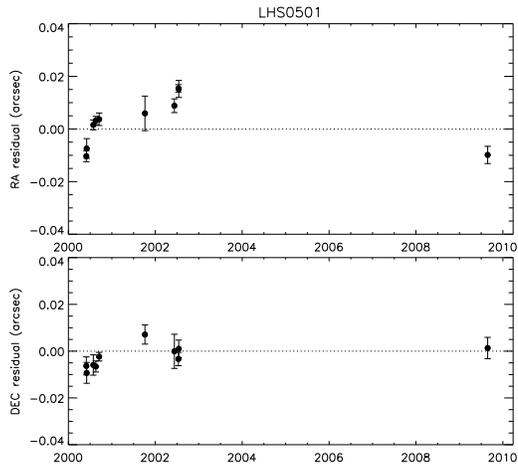}
  \label{fig:lhs0501A}}
  \caption{Nightly mean astrometric residuals in right ascension and
  declination are shown for LHS 440, LHS 327 (a typical star without a
  detected perturbation), and LHS 501.  The astrometric signatures of
  each system's proper motion and parallax have been removed. [To
  Editor: Please arrange three plots horizontally from left (a) to
  right (c).]}
  \label{fig:perturb}
  \end{figure}

{\bf LHS 440} As shown in Figure~\ref{fig:perturb}, this star shows a
convincing perturbation in the residuals on the DEC axis that is
possibly confirmed in the RA axis.  The period is longer than 10
years, so the star will remain on CTIOPI because it is subdwarf of
type M1.0VI \citep{Jao2008}, making it a rare example of a nearby
subdwarf binary that may eventually yield important dynamical masses.

{\bf LHS 499} Images taken during the first season used the old $V$
filter, while the rest of three seasons were observed using new $V$.
Our $\pi_{trig}=$ 56.93$\pm$2.12 mas uses data from both filters and
is consistent with the YPC value of 63.0$\pm$11.7 mas with a smaller
error by a factor of 5.5.

{\bf LHS 501 and LHS 500} This is a wide common proper motion binary
with separation 107\farcs1 at position angle 185.2$^{\circ}$
\citep{Jao2003}.  LHS 501, the primary star A, shows a perturbation on
both axes, as shown in Figure~\ref{fig:perturb}.  We do not yet have
sufficient data to derive an orbit, but the system will remain on
CTIOPI.  The perturbation is likely the cause of the discrepant
parallaxes and proper motions determined for A and B, where the values
for the B component are more reliable.  Preliminary high resolution
imaging using the Gemini-North adaptive optics system (ALTAIR) and
Near InfraRed Imager (NIRI) in 2009 has resolved the tertiary near LHS
501.  Details about these AO results will be discussed in a future
paper.

{\bf LHS 521} This star shows a possible perturbation in DEC that does
not appear in the series for any reference star.  However, more
observations are needed to confirm the perturbation.

{\bf GJ 1277} Our relative parallax measurement of 96.31$\pm$0.95 mas
is consistent with the relative parallax measurement of 94$\pm$1.0 mas
reported by \citet{Bartlett2009}, who did not provide an absolute
parallax.

\section{Summary}
\label{discussion}

We present 41 new and revised trigonometric parallaxes of 37 systems,
of which 15 are red dwarfs and 22 are subdwarfs. By combining the
results of this paper, \citet{Jao2005}, \citet{Costa2005}, and
\citet{Costa2006} (the latter two papers from our CTIO 1.5-m parallax
program), we have now determined parallaxes for 32 subdwarf systems.
This comprises the largest set of cool subdwarf parallaxes since
\citet{Monet1992}'s effort that included 21 cool subdwarf systems.
Our continuing efforts will further increase the number of nearby
subdwarfs with accurate parallaxes (errors less than 10\%), and will
allow us to better characterize the population of cool subdwarfs in
the solar neighborhood.

Unlike dwarfs, subdwarfs show very different spectroscopic ``flavors''
depending on their metallicities and gravities \citep{Jao2008}.  This
is evident in Figure~\ref{fig:color.mag.1}, where the spread in
$M_{Ks}$ is greater than three magnitudes for some $(V-K_s)$ colors.
Thus, building useful combinations of $VRIJHK_{s}$ photometry to
estimate distances to subdwarfs is far more difficult than for dwarfs,
which is relatively straightforward \citep{Henry2004}.  In fact, if
one wrongly applies relations for main-sequence dwarfs or incorrect
subdwarf ``flavors'' to a specific subdwarf, it will be estimated to
be either further away or much closer that it truly is.  By measuring
more trigonometric parallaxes and accurate photometry for cool
subdwarfs, we can develop better distance estimating techniques and
perhaps reveal more nearby subdwarfs like LHS 272.

In the process of investigating the present sample of high proper
motion stars and subdwarfs, we have detected a rare close subdwarf
binary (LHS 440) as well as a dwarf binary (LHS 501) via astrometric
perturbations.  The only cool subdwarf with an accurate mass
measurement is $\mu$ Cas B \citep[0.17 M$_{\odot}$,][]{McCarthy1993}.
Our continued astrometric efforts coupled with future radial velocity
observations will yield essential mass measurements for low-mass
subdwarfs, and thus provide important dynamical mass points on the
subdwarf empirical mass-luminosity relation.

Finally, we have used our long term parallax observations to compare
the photometric variability of cool dwarfs and subdwarfs.  Subdwarfs
appear to be less variable than dwarfs in the $R$ band, where we have
the most data points, and possibly less variable than dwarfs in the
$V$ band.  However, our variability techniques should be applied to
much larger subdwarf samples to confirm these intriguing results.

\section{Acknowledgments}

The astrometric observations reported here began as part of the NOAO
Surveys Program in 1999 and continued on the CTIO 0.9-m via the SMARTS
Consortium starting in 2003.  We gratefully acknowledge support from
the National Science Foundation (grants AST 05-07711 and AST
09-08402), NASA's Space Interferometry Mission, and Georgia State
University, which together have made this long-term effort possible.

We thank the referee to provide comments to improve this paper. We
thank Po-Yung Chen for his efforts on statistics. We also thank the
members of the SMARTS Consortium, who enable the operations of the
small telescopes at CTIO, as well as the supporting observers support
at CTIO, specifically Edgardo Cosgrove, Arturo G\'{o}mez, Alberto
Miranda, and Joselino V\'{a}squez. This research has made use of the
SIMBAD database, operated at CDS, Strasbourg, France. This work also
has used data products from the Two Micron All Sky Survey, which is a
joint project of the University of Massachusetts and the Infrared
Processing and Analysis Center at California Institute of Technology
funded by NASA and NSF.




  
\begin{deluxetable}{lccccccccrrrrrrc}
\rotate
\setlength{\tabcolsep}{0.03in}
\tablewidth{0pt}
\tabletypesize{\tiny}
\tablehead{\colhead{Name}                &
	   \colhead{RA}                  &
 	   \colhead{DEC}                 &
 	   \colhead{Filt}                &
	   \colhead{Nsea}                &
	   \colhead{Nfrm}                &
	   \colhead{Coverage}            &
	   \colhead{Years}               &
	   \colhead{Nref}                &
	   \colhead{$\pi$(rel)}          &
	   \colhead{$\pi$(corr)}         &
	   \colhead{$\pi$(abs)}          &
	   \colhead{$\mu$}               &
	   \colhead{P.A.}                &
	   \colhead{$V_{tan}$}           &
	   \colhead{Note}                \\
	   \colhead{}                    & 
	   \multicolumn{2}{c}{(J2000.0)} &
           \colhead{}                    &
	   \colhead{}                    &
	   \colhead{}                    &
	   \colhead{}                    &
	   \colhead{}                    &
	   \colhead{}                    &
	   \colhead{(mas)}               &
	   \colhead{(mas)}               &
	   \colhead{(mas)}               &
	   \colhead{(mas/yr)}            &
	   \colhead{(deg)}               &
	   \colhead{(km/s)}              &
	   \colhead{}                    \\
           \colhead{(1)}                 &
           \multicolumn{2}{c}{(2)}       &
           \colhead{(3)}                 &
           \colhead{(4)}                 &
           \colhead{(5)}                 &
           \colhead{(6)}                 &
           \colhead{(7)}                 &
           \colhead{(8)}                 &
           \colhead{(9)}                 &
           \colhead{(10)}                &
           \colhead{(11)}                &
           \colhead{(12)}                &
           \colhead{(13)}                &
           \colhead{(14)}                }
\startdata
\multicolumn{15}{c}{First Trigonometric Parallaxes}\\				   
\hline

G 266-089B                     &   00 19 36.59 & $-$28 09 38.8 &       V  &     5s &  62 &  2000.87--2005.69 &  4.82 &   8  &     33.26$\pm$1.68  & 1.39$\pm$0.14  &  34.65$\pm$1.69  &  1370.8$\pm$0.9  & 192.1$\pm$0.07  & 187.5  &   !   \\ 
G 266-089A                     &   00 19 37.02 & $-$28 09 45.7 &       V  &     5s &  62 &  2000.87--2005.69 &  4.82 &   8  &     31.45$\pm$1.68  & 1.39$\pm$0.14  &  32.84$\pm$1.69  &  1372.2$\pm$0.9  & 192.3$\pm$0.07  & 198.1  &   !   \\ 
LHS 124                        &   00 49 29.05 & $-$61 02 32.7 &       V  &     9s &  57 &  2000.88--2008.87 &  7.99 &   7  &     47.44$\pm$1.38  & 1.18$\pm$0.14  &  48.62$\pm$1.39  &  1126.5$\pm$0.6  &  94.6$\pm$0.05  & 109.8  &       \\ 
LHS 125                        &   00 50 17.09 & $-$39 30 08.3 &       R  &     7s &  48 &  2001.88--2008.86 &  6.98 &   5  &     11.90$\pm$3.49  & 2.06$\pm$0.16  &  13.96$\pm$3.49  &  1031.5$\pm$1.3  & 171.4$\pm$0.12  & 350.3  &       \\ 
LHS 164                        &   03 01 40.58 & $-$34 57 56.5 &       R  &     8s &  83 &  2001.87--2009.03 &  7.16 &   7  &     18.19$\pm$1.55  & 0.97$\pm$0.11  &  19.16$\pm$1.55  &  1323.9$\pm$0.7  & 157.7$\pm$0.05  & 327.5  &       \\ 
LHS 176                        &   03 35 38.61 & $-$08 29 22.7 &       I  &     5s &  47 &  2003.95--2009.12 &  5.17 &   7  &     76.35$\pm$1.30  & 1.42$\pm$0.08  &  77.77$\pm$1.30  &  1549.8$\pm$0.7  & 101.9$\pm$0.04  &  94.5  &       \\ 
SCR0342-6407                   &   03 42 57.40 & $-$64 07 56.5 &       I  &     4s &  66 &  2003.94--2007.89 &  3.95 &   9  &     41.13$\pm$2.01  & 0.43$\pm$0.04  &  41.56$\pm$2.01  &  1059.9$\pm$0.9  & 143.3$\pm$0.10  & 120.9  &       \\ 
WT0135                         &   04 11 27.14 & $-$44 18 09.7 &       R  &     7s &  59 &  2000.07--2009.78 &  9.71 &   5  &     38.38$\pm$2.42  & 0.66$\pm$0.04  &  39.04$\pm$2.42  &   691.5$\pm$0.8  &  67.1$\pm$0.13  &  84.0  &       \\
LSR0627+0616                   &   06 27 33.33 & $+$06 16 58.9 &       I  &     4c+&  41 &  2002.95--2009.02 &  6.07 &  14  &     14.68$\pm$1.21  & 1.75$\pm$0.32  &  16.43$\pm$1.25  &  1009.0$\pm$0.9  & 179.0$\pm$0.07  & 291.1  &   !   \\ 
LHS 272                        &   09 43 46.16 & $-$17 47 06.2 &       V  &     4s &  61 &  2001.15--2005.06 &  3.91 &  10  &     72.81$\pm$1.17  & 1.14$\pm$0.11  &  73.95$\pm$1.18  &  1439.0$\pm$1.0  & 279.2$\pm$0.07  &  92.2  &   !   \\ 
LHS 284                        &   10 36 03.09 & $-$14 42 29.1 &       I  &     3c &  58 &  2003.08--2005.21 &  2.13 &  12  &     20.35$\pm$1.30  & 0.79$\pm$0.07  &  21.14$\pm$1.30  &  1085.6$\pm$1.3  & 297.8$\pm$0.13  & 243.4  &       \\ 
SCR1107-4135                   &   11 07 55.90 & $-$41 35 52.8 &       I  &     4s &  55 &  2006.21--2009.25 &  3.04 &   8  &     13.82$\pm$1.18  & 0.97$\pm$0.08  &  14.79$\pm$1.18  &  1186.5$\pm$1.0  & 283.2$\pm$0.08  & 380.1  &       \\ 
LHS 323                        &   12 17 30.16 & $-$29 02 20.7 &       I  &     4s &  46 &  2006.21--2009.23 &  3.02 &   8  &     22.46$\pm$1.83  & 0.83$\pm$0.04  &  23.29$\pm$1.83  &  1105.8$\pm$1.4  & 267.1$\pm$0.11  & 225.0  &       \\ 
LHS 327                        &   12 25 50.73 & $-$24 33 17.8 &       R  &     6s &  50 &  2001.15--2009.31 &  8.16 &   9  &      9.98$\pm$1.40  & 0.84$\pm$0.11  &  10.72$\pm$1.40  &   981.6$\pm$0.6  & 262.6$\pm$0.05  & 433.8  &   !   \\ 
GJ 1158                        &   12 29 34.54 & $-$55 59 37.1 &       V  &     6s+&  76 &  2001.15--2008.21 &  7.06 &  10  &     73.60$\pm$1.15  & 2.58$\pm$0.77  &  76.18$\pm$1.38  &  1205.0$\pm$0.5  & 228.0$\pm$0.05  &  75.0  &       \\ 
LHS 347                        &   13 10 01.80 & $+$22 30 05.3 &       R  &     6s &  54 &  2001.15--2009.32 &  8.17 &   5  &     15.73$\pm$2.29  & 0.99$\pm$0.09  &  16.72$\pm$2.29  &  1169.8$\pm$1.0  & 232.3$\pm$0.09  & 331.6  &   !   \\ 
LHS 2734A                      &   13 25 14.20 & $-$21 27 12.4 &       I  &     3s &  36 &  2003.09--2005.48 &  2.39 &  12  &      2.79$\pm$1.19  & 1.15$\pm$0.09  &   3.94$\pm$1.19  &   594.2$\pm$1.1  & 227.8$\pm$0.21  & 714.8  &   !   \\ 
LHS 2734B                      &   13 25 15.70 & $-$21 28 18.0 &       I  &     3s &  36 &  2003.09--2005.48 &  2.39 &  12  &   $-$1.70$\pm$1.52  & 1.15$\pm$0.09  &   3.94$\pm$1.19* &   595.2$\pm$1.4  & 228.2$\pm$0.27  &\nodata &   !   \\ 
LHS 3045                       &   15 14 54.39 & $-$31 50 13.6 &       R  &     5s+&  51 &  2006.21--2009.61 &  3.40 &   8  &     14.61$\pm$0.98  & 2.38$\pm$0.29  &  16.99$\pm$1.02  &   930.7$\pm$0.9  & 217.8$\pm$0.11  & 259.7  &   !   \\
SIPS1529-2907                  &   15 29 14.00 & $-$29 07 37.7 &       I  &     4s &  45 &  2006.22--2009.31 &  3.09 &  11  &     26.64$\pm$1.02  & 1.18$\pm$0.17  &  27.82$\pm$1.03  &  1016.9$\pm$0.9  & 187.7$\pm$0.08  & 187.7  &       \\ 
SCR1916-3638                   &   19 16 46.57 & $-$36 38 05.9 &       I  &     6s &  56 &  2005.72--2009.49 &  3.77 &   8  &     12.51$\pm$1.19  & 2.27$\pm$0.68  &  14.78$\pm$1.37  &  1293.5$\pm$0.9  & 184.5$\pm$0.06  & 414.8  &       \\ 
LHS 3620                       &   21 04 25.37 & $-$27 52 46.8 &       I  &     6s &  49 &  2003.52--2009.57 &  6.05 &   9  &     11.74$\pm$1.40  & 1.12$\pm$0.11  &  12.88$\pm$1.40  &   968.7$\pm$0.9  & 186.4$\pm$0.09  & 357.1  &       \\ 
SCR 2115-7541                  &   21 15 15.09 & $-$75 41 52.0 &       I  &     4s &  75 &  2003.51--2006.57 &  3.06 &  12  &     30.88$\pm$1.26  & 2.05$\pm$0.22  &  32.96$\pm$1.28  &  1052.7$\pm$1.0  & 144.6$\pm$0.10  & 151.5  &       \\ 
LHS 3732                       &   21 55 57.10 & $-$45 39 34.3 &       R  &     4c &  40 &  2005.70--2008.64 &  3.14 &   9  &      6.59$\pm$1.57  & 1.75$\pm$0.18  &   8.34$\pm$1.58  &   959.3$\pm$1.6  & 157.9$\pm$0.18  & 545.1  &   !   \\ 
LHS 3740                       &   21 58 53.18 & $-$57 56 03.5 &       R  &     5s &  52 &  2005.72--2009.78 &  4.06 &   5  &     34.74$\pm$1.45  & 1.20$\pm$0.10  &  35.94$\pm$1.45  &   908.9$\pm$1.0  &  95.7$\pm$0.09  & 121.8  &       \\ 
LHS 518                        &   22 20 26.97 & $-$24 21 49.5 &       R  &     5s+&  84 &  2002.51--2009.54 &  7.03 &  11  &     15.41$\pm$1.31  & 0.51$\pm$0.09  &  15.92$\pm$1.31  &  1063.9$\pm$0.7  & 156.0$\pm$0.07  & 316.8  &       \\ 
\hline											                     
\multicolumn{15}{c}{Revised Parallaxes}\\				   		                     
\hline											                     
LHS 193B                       &   04 32 35.98 & $-$39 02 14.6 &       V  &     6s &  78 &  2000.87--2009.74 &  8.87 &   8  &     30.77$\pm$2.40  & 2.21$\pm$0.21  &  32.98$\pm$2.41  &   999.2$\pm$1.2  &  44.7$\pm$0.14  & 143.6  &       \\
LHS 193A                       &   04 32 36.56 & $-$39 02 03.4 &       V  &     6s &  78 &  2000.87--2009.74 &  8.87 &   8  &     32.55$\pm$1.43  & 2.21$\pm$0.21  &  34.76$\pm$1.45  &   993.8$\pm$0.6  &  44.7$\pm$0.07  & 135.5  &  1    \\
LHS 205                        &   05 16 59.67 & $-$78 17 20.2 &       V  &     7s &  64 &  2003.95--2009.75 &  5.80 &   6  &     64.48$\pm$1.89  & 0.87$\pm$0.24  &  65.35$\pm$1.91  &  1134.1$\pm$1.2  & 179.0$\pm$0.09  &  82.3  &  2    \\
GJ 1129                        &   09 44 47.34 & $-$18 12 48.9 &       V  &     5s &  48 &  2000.06--2004.33 &  4.27 &   7  &     92.38$\pm$2.48  & 1.51$\pm$0.30  &  93.89$\pm$2.49  &  1597.3$\pm$1.3  & 264.3$\pm$0.07  &  80.6  &  3    \\
LHS 300AB                      &   11 11 13.68 & $-$41 05 32.7 &       R  &     6s &  76 &  2001.15--2009.31 &  8.16 &  11  &     31.38$\pm$1.35  & 1.65$\pm$0.17  &  33.03$\pm$1.36  &  1251.5$\pm$0.6  & 263.9$\pm$0.04  & 179.6  &  4    \\
LHS 318                        &   11 56 54.87 & $+$26 39 56.3 &       I  &     5s &  57 &  2003.25--2009.25 &  6.00 &   6  &     18.10$\pm$2.32  & 0.66$\pm$0.05  &  18.76$\pm$2.32  &  1360.3$\pm$1.4  & 154.5$\pm$0.11  & 343.7  &  5, ! \\ 
LHS 326                        &   12 24 26.81 & $-$04 43 36.7 &       R  &     4s &  66 &  2003.09--2008.38 &  5.29 &   7  &     19.59$\pm$1.94  & 0.80$\pm$0.07  &  20.39$\pm$1.94  &  1301.3$\pm$0.7  & 241.8$\pm$0.06  & 302.4  &  6, ! \\ 
LHS 406                        &   15 43 18.33 & $-$20 15 32.9 &       R  &     6s &  81 &  2000.57--2009.31 &  8.74 &  12  &     44.98$\pm$1.14  & 1.75$\pm$0.28  &  46.73$\pm$1.17  &  1160.8$\pm$0.5  & 194.8$\pm$0.04  & 117.7  &  7    \\
LHS 440                        &   17 18 25.58 & $-$43 26 37.6 &       R  &     6s & 100 &  2000.58--2009.58 &  9.00 &  10  &     34.52$\pm$1.09  & 1.88$\pm$0.54  &  36.40$\pm$1.22  &  1082.5$\pm$0.5  & 233.3$\pm$0.05  & 141.0  &  8, ! \\
LHS 475                        &   19 20 54.26 & $-$82 33 16.1 &       V  &     9s & 132 &  2000.57--2009.54 &  8.97 &   8  &     81.68$\pm$0.93  & 1.36$\pm$0.08  &  83.04$\pm$0.93  &  1269.6$\pm$0.3  & 164.5$\pm$0.02  &  72.5  &  9    \\
LHS 499                        &   20 51 41.64 & $-$79 18 39.9 &       V  &     4s &  63 &  2004.56--2007.75 &  3.19 &   7  &     55.51$\pm$2.11  & 1.42$\pm$0.17  &  56.93$\pm$2.12  &  1209.2$\pm$1.7  & 143.9$\pm$0.16  & 100.7  & 10, ! \\ 
LHS 500                        &   20 55 37.12 & $-$14 03 54.8 &       V  &     5s &  71 &  1999.70--2009.73 & 10.03 &   9  &     81.91$\pm$1.24  & 0.88$\pm$0.05  &  82.79$\pm$1.24  &  1490.8$\pm$0.5  & 108.3$\pm$0.03  &  85.4  & 11, ! \\
LHS 501                        &   20 55 37.76 & $-$14 02 08.1 &       V  &     5s &  71 &  1999.70--2009.73 & 10.03 &   9  &     72.20$\pm$1.17  & 0.88$\pm$0.05  &  73.08$\pm$1.17  &  1297.6$\pm$0.4  & 108.2$\pm$0.03  &  84.2  & 12, ! \\
LHS 521                        &   22 27 59.21 & $-$30 09 32.8 &       R  &     5c &  76 &  2000.58--2009.54 &  8.96 &   8  &     17.24$\pm$1.07  & 1.22$\pm$0.09  &  18.46$\pm$1.07  &  1006.8$\pm$0.6  & 137.2$\pm$0.07  & 258.5  & 13, ! \\
GJ 1277                        &   22 56 24.66 & $-$60 03 49.2 &       V  &     8s+&  80 &  2001.87--2007.82 &  5.95 &   7  &     96.31$\pm$0.95  & 1.17$\pm$0.69  &  97.48$\pm$1.17   & 1082.0$\pm$0.6  & 210.4$\pm$0.06  &  52.6  & 14, ! \\

\enddata

\tablecomments{ N$_{sea}$ indicates the number of seasons observed,
  where 2--3 months of observations count as one season, for seasons
  having more than 3 images taken.  The letter ``c'' indicates a
  continuous set of observations where multiple nights of data were
  taken in each season, whereas an ``s'' indicates scattered
  observations when one or more seasons have only a single night of
  observations.  Generally, ``c'' observations are better.  A $+$
  indicates that three or fewer individual images are used in one or
  more seasons that are not counted in N$_{sea}$.  Stars with
  exclamation marks in the Notes column are discussed in
  Section~\ref{sec:notes}.  The * indicates that the absolute parallax
  of LHS 2734B has been adopted from LHS 2734A.  All previous parallax
  measurements listed here are absolute parallaxes, other than GJ1277,
  which is a relative parallax: (1) Parallax of 32.06$\pm$1.65 mas in
  \citet{Jao2005}. (2) Parallax of 77.50$\pm$11 mas in YPC. (3)
  Parallax of 90.93$\pm$3.78 mas in \citet{Jao2005}. (4) Parallax of
  32.30$\pm$1.85 mas in \citet{Jao2005}. (5) Parallax of 24.8$\pm$6
  mas in \citet{Smart2007}. (6) Parallax of 11.7$\pm$4.3 mas in
  \citet{Smart2007}. (7) Parallax of 47.28$\pm$1.61 mas in
  \citet{Jao2005}. (8) Parallax of 36.90$\pm$2.19 mas in
  \citet{Jao2005}. (9) Parallax of 78.34$\pm$2.03 mas in
  \citet{Jao2005}. (10) Parallax of 63.0$\pm$11.7 mas in YPC. (11)
  Parallax of 81.95$\pm$1.54 mas in \citet{Jao2005}. (12) Parallax of
  77.59$\pm$1.49 mas in \citet{Jao2005}. (13) Parallax of
  21.60$\pm$1.59 mas in \citet{Jao2005}. (14) Relative Parallax of
  94$\pm$1.0 mas in \citet{Bartlett2009}.}

\label{tbl:pi.result}
\end{deluxetable}


\begin{deluxetable}{llrrrccccrrrlc}
\rotate
\tablewidth{0pt}
\tabletypesize{\tiny}
\tablehead{
           \colhead{Name1}   &
           \colhead{Name2}   &
           \colhead{$V$}     &
           \colhead{$R$}     &
           \colhead{$I$}     &
           \colhead{\#}      &
	   \colhead{$\pi$}   &
	   \colhead{$\sigma$}&
           \colhead{No. of Nights}    &
           \colhead{$J$}     &
           \colhead{$H$}     &
           \colhead{$K_{s}$} &
	   \colhead{Spect.}  &
	   \colhead{Refs}    \\
	   \colhead{}        &
	   \colhead{}        &
	   \colhead{mag}     &
	   \colhead{mag}     &
	   \colhead{mag}     &
	   \colhead{}        &
	   \colhead{filter}  &
	   \colhead{mag}     &
	   \colhead{}        &
	   \colhead{mag}     &
	   \colhead{mag}     &
	   \colhead{mag}     &
	   \colhead{}        &
	   \colhead{}        \\
	   \colhead{(1)}     &
	   \colhead{(2)}     &
	   \colhead{(3)}     &
	   \colhead{(4)}     &
	   \colhead{(5)}     &
	   \colhead{(6)}     &
	   \colhead{(7)}     &
           \colhead{(8)}     &
           \colhead{(9)}     &
           \colhead{(10)}    &
	   \colhead{(11)}    &
	   \colhead{(12)}    & 
	   \colhead{(13)}    & 
	   \colhead{(14)}     
           }
\startdata
  G 266-089B      &  LHS 110      &  14.19 &    13.17 &    11.87 &    2 & V  &  0.008 & 13 &     10.63$\pm$0.02        &   10.12$\pm$0.03       &  9.88$\pm$0.02     & M4.0J[VI]  & 1        \\
  G 266-089A      &  LHS 111      &  13.65 &    12.66 &    11.44 &    2 & V  &  0.007 & 13 &     10.25$\pm$0.02        &    9.73$\pm$0.03       &  9.48$\pm$0.02     & M4.0J[VI]  & 1        \\
  LHS 124         &  GJ 1022      &  12.17 &    11.12 &     9.86 &    2 & V  &  0.011 & 15 &      8.63$\pm$0.02        &    8.09$\pm$0.05       &  7.84$\pm$0.03     & M2.0V      & 1        \\
  LHS 125         &  LP 989-183   &  14.32 &    13.58 &    12.92 &    2 & R  &  0.010 & 10 &     12.05$\pm$0.02        &   11.58$\pm$0.02       & 11.45$\pm$0.03     & K4.0[VI]   & 6        \\
  LHS 164         &  LEHPM 2991   &  13.56 &    12.81 &    12.14 &    3 & R  &  0.007 & 18 &     11.38$\pm$0.03        &   10.77$\pm$0.02       & 10.64$\pm$0.02     & K7.0VI     & 6        \\
  LHS 176         &  NLTT 11328   &  15.92 &    14.30 &    12.31 &    3 & I  &  0.009 & 11 &     10.38$\pm$0.02        &    9.80$\pm$0.02       &  9.46$\pm$0.02     & M5.0V      & 1        \\
  SCR 0342-6407   &               &  16.01 &    14.65 &    12.89 &    2 & I  &  0.006 & 13 &     11.32$\pm$0.02        &   10.87$\pm$0.03       & 10.58$\pm$0.02     & M4.0V      & 1        \\
  WT0135          &  LEHPM 3673   &  14.10 &    13.06 &    11.82 &    2 & R  &  0.007 & 12 &     10.55$\pm$0.02        &   10.12$\pm$0.03       &  9.83$\pm$0.02     & M3.0VI     & 6        \\
  LSR 0627+0616   &               &  16.28 &    15.31 &    14.37 &    2 & I  &  0.008 & 10 &     13.29$\pm$0.03        &   12.83$\pm$0.03       & 12.63$\pm$0.03     & esdM1.5    & 7        \\
  LHS 272         &  NLTT 22460   &  13.16 &    12.10 &    10.87 &    3 & V  &  0.015 & 13 &      9.62$\pm$0.02        &    9.12$\pm$0.02       &  8.87$\pm$0.02     & M3.0VI     & 6        \\
  LHS 284         &  NLTT 24803   &  16.78 &    15.49 &    13.81 &    3 & I  &  0.007 & 13 &     12.28$\pm$0.02        &   11.79$\pm$0.03       & 11.58$\pm$0.03     & M4.0V      & 1        \\
  SCR1107-4135    &               &  14.96 &    14.07 &    13.21 &    2 & I  &  0.006 & 12 &     12.19$\pm$0.02        &   11.69$\pm$0.02       & 11.47$\pm$0.02     & M0.5VI     & 6        \\
  LHS 323         &  NLTT 30238   &  16.95 &    15.66 &    14.02 &    2 & I  &  0.008 &  9 &     12.54$\pm$0.02        &   12.05$\pm$0.02       & 11.78$\pm$0.02     & M4.0V      & 1        \\
  LHS 327         &  NLTT 30709   &  12.75 &    12.17 &    11.62 &    2 & R  &  0.010 & 10 &     10.81$\pm$0.02        &   10.31$\pm$0.02       & 10.14$\pm$0.02     & K4.0[VI]   & 6        \\
  GJ1158          &  LHS 322      &  13.26 &    12.02 &    10.41 &    3 & V  &  0.014 & 16 &      8.89$\pm$0.03        &    8.35$\pm$0.04       &  8.07$\pm$0.02     & M3.0V      & 2        \\
  LHS 347         &  NLTT 33109   &  12.42 &    11.75 &    11.13 &    2 & R  &  0.007 & 11 &     10.26$\pm$0.02        &    9.71$\pm$0.03       &  9.59$\pm$0.02     & K2.0[VI]   & 1        \\
  LHS 2734A       &  LP 797-61    &  16.13 &    15.32 &    14.59 &    2 & I  &  0.005 &  7 &     13.63$\pm$0.03        &   13.10$\pm$0.03       & 12.90$\pm$0.04     & K7.0VI     & 6        \\
  LHS 2734B       &               &  18.79 &    17.93 &    16.79 &    2 & I  &  0.020 &  7 &     15.83$\pm$0.09        &   15.30$\pm$0.09       & 14.93$\pm$0.14     & M1.0VI:    & 6        \\
  LHS 3045        &  NLTT 39664   &  14.39 &    13.54 &    12.74 &    2 & R  &  0.007 & 13 &     11.76$\pm$0.02        &   11.23$\pm$0.03       & 11.00$\pm$0.02     & [VI]       & 1        \\
  SIPS 1529-2907  &               &  19.38 &    17.52 &    15.28 &    4 & I  &  0.008 & 11 &     13.32$\pm$0.03        &   12.86$\pm$0.02       & 12.50$\pm$0.02     & M6.5V      & 1        \\
  SCR 1916-3638   &               &  16.83 &    15.82 &    14.76 &    3 & I  &  0.008 & 13 &     13.66$\pm$0.02        &   13.12$\pm$0.02       & 12.95$\pm$0.03     & M3.0VI     & 6        \\
  LHS 3620        &  NLTT 50449   &  16.61 &    15.59 &    14.53 &    2 & I  &  0.007 & 13 &     13.41$\pm$0.02        &   12.89$\pm$0.02       & 12.70$\pm$0.03     & M2.0VI     & 6        \\
  SCR 2115-7541   &               &  14.48 &    13.25 &    11.66 &    3 & I  &  0.009 & 15 &     10.14$\pm$0.02        &    9.60$\pm$0.02       &  9.33$\pm$0.02     & M3.5V      & 1        \\
  LHS 3732        &  L 355-29     &  14.11 &    13.35 &    12.64 &    3 & R  &  0.005 &  9 &     11.72$\pm$0.02        &   11.14$\pm$0.02       & 11.00$\pm$0.02     & VI         & 4        \\
  LHS 3740        &  L 213-75     &  14.06 &    12.88 &    11.33 &    2 & R  &  0.009 & 13 &      9.85$\pm$0.02        &    9.27$\pm$0.02       &  9.00$\pm$0.02     & M3.5V      & 1        \\
  LHS 518         &  NLTT 53550   &  13.63 &    12.80 &    12.01 &    2 & R  &  0.008 & 18 &     11.04$\pm$0.02        &   10.59$\pm$0.02       & 10.39$\pm$0.02     & M1.0VI     & 6        \\
\hline										     												       
  LHS 193B        &               &  17.73 &    17.18 &    16.60 &    3 & V  &  0.049 & 13 &     16.21$\pm$\nodata     &   15.94$\pm$\nodata    &\nodata$\pm$\nodata & WD         & 5        \\
  LHS 193A        &  L 447-10     &  11.66 &    10.85 &    10.09 &    3 & V  &  0.005 & 13 &      9.18$\pm$0.02        &    8.55$\pm$0.02       &  8.43$\pm$0.02     & K6.0VI     & 6        \\
  LHS 205         &  GJ 1077      &  11.90 &    10.81 &     9.42 &    4 & V  &  0.007 & 14 &      8.07$\pm$0.02        &    7.44$\pm$0.02       &  7.20$\pm$0.02     & M2.0V      & 2        \\
  GJ 1129         &  LHS 273      &  12.46 &    11.24 &     9.67 &    3 & V  &  0.026 & 10 &      8.12$\pm$0.03        &    7.54$\pm$0.04       &  7.26$\pm$0.02     & M3.5V      & 3        \\
  LHS 300A        &  L 395-13     &  13.18 &    12.28 &    11.49 &    2 & R  &  0.008 & 13 &     10.48$\pm$0.02        &   10.01$\pm$0.03       &  9.80$\pm$0.02     & M0.0VI:    & 6        \\
  LHS 318         &  NLTT 29045   &  15.41 &    14.48 &    13.53 &    3 & I  &  0.006 & 14 &     12.50$\pm$0.02        &   11.98$\pm$0.02       & 11.80$\pm$0.02     & M2.0VI:    & 6        \\
  LHS 326         &  NLTT 30636   &  14.92 &    13.98 &    13.04 &    2 & R  &  0.008 & 14 &     11.93$\pm$0.02        &   11.43$\pm$0.02       & 11.23$\pm$0.02     & M3.0VI     & 6        \\
  LHS 406         &  NLTT 40994   &  13.06 &    12.07 &    10.93 &    2 & R  &  0.009 & 17 &      9.78$\pm$0.02        &    9.23$\pm$0.02       &  9.02$\pm$0.02     & M2.0VI     & 6        \\
  LHS 440         &  L 413-156    &  12.98 &    11.98 &    10.86 &    2 & R  &  0.008 & 17 &      9.70$\pm$0.02        &    9.13$\pm$0.02       &  8.95$\pm$0.02     & M1.0VI:    & 6        \\
  LHS 475         &  L 22-69      &  12.69 &    11.51 &    10.00 &    4 & V  &  0.011 & 26 &      8.56$\pm$0.03        &    8.00$\pm$0.04       &  7.69$\pm$0.04     & M3.0V      & 2        \\
  LHS 499         &  GJ 808       &  11.81 &    10.82 &     9.64 &    2 & V  &  0.006 & 12 &      8.46$\pm$0.03        &    7.91$\pm$0.05       &  7.66$\pm$0.02     & M1.5V      & 2        \\
  LHS 500         &  GJ 810B      &  14.63 &    13.21 &    11.40 &    2 & V  &  0.014 & 14 &      9.72$\pm$0.02        &    9.22$\pm$0.02       &  8.92$\pm$0.02     & M5.0V      & 2        \\
  LHS 501         &  GJ 810A      &  12.48 &    11.23 &     9.62 &    2 & V  &  0.014 & 14 &      8.12$\pm$0.03        &    7.64$\pm$0.04       &  7.37$\pm$0.03     & M4.0V      & 2        \\
  LHS 521         &  LP 932-1     &  14.70 &    13.85 &    13.10 &    2 & R  &  0.008 & 16 &     12.13$\pm$0.02        &   11.66$\pm$0.03       & 11.46$\pm$0.02     & M0.5VI:    & 6        \\
  GJ1277          &  LHS 532      &  14.00 &    12.59 &    10.79 &    3 & V  &  0.006 & 20 &      8.98$\pm$0.03        &    8.36$\pm$0.03       &  8.11$\pm$0.02     & M4.5V      & 1        \\
 \enddata

\tablecomments{Brackets [~] indicate a possible luminosity class. A
  colon : indicates a questionable subtype.}

\tablerefs{(1) this work; (2) \citealt{Hawley1996}; (3)
  \citealt{Henry2002}; (4) \citealt{Ibata1997}; (5) \citealt{Jao2005};
  (6) \citealt{Jao2008}; (7) \citealt{Lepine2003}}

\label{tbl:phot.result}
\end{deluxetable}


\begin{table}
\begin{tabular}{lccccc}
\tableline
\tableline
Name        &  \multicolumn{2}{c}{Old $V$} &  \multicolumn{2}{c}{New $V$}  &  Combined\\
            &  $\sigma_{mag}$&  Nfrm    & $\sigma_{mag}$&  Nfrm   & $\sigma_{mag}$  \\

\tableline
LHS 145     &  0.007 & 115      &  0.005 & 50  &  0.007  \\     
GJ 440      &  0.008 & 124      &  0.006 & 50  &  0.007  \\
GJ 781.3    &  0.007 & 86       &  0.006 & 48  &  0.007  \\
\tableline
\end{tabular}
\caption{Variability comparisons of three white dwarfs using the old
  and new V filters.  The $\sigma_{mag}$ values represent the standard
  deviations of the photometry calculated via the method described in
  the text.  The ``Nfrm'' values correspond to the total number of
  frames used in each filter.  The combined variabilities are from
  \cite{Subasavage2009} and have been measured using the same
  methodology.}
\label{tbl:varia.wd}
\end{table}

\end{document}